\documentclass[conference]{IEEEtran}
\IEEEoverridecommandlockouts
\usepackage{amsthm}
\usepackage{amsmath}
\usepackage{amssymb}
\usepackage{booktabs}
\usepackage{siunitx}
\usepackage{cite}
\usepackage{epsfig}
\usepackage{epstopdf}
\usepackage{url}
\usepackage{multimedia}
\usepackage{paralist}
\usepackage[printonlyused]{acronym}
\usepackage{units}

\newcommand{\SEC}[1]{Section~\ref{#1}}

\def\pathloss{\ensuremath{n}}

\acrodef{5G}[5G]{5\textsuperscript{th}--Generation}
\acrodef{BW}[BW]{bandwidth}
\acrodef{CW}[CW]{continuous wave}
\acrodef{D2D}[D2D]{device--to--device}
\acrodef{dB}[dB]{decibel}
\acrodef{dBi}[dBi]{decibel isotropic}
\acrodef{dBm}[dBm]{decibel over a milliwatt}
\acrodef{Gbps}[Gbps]{Gigabit per second}
\acrodef{GHz}[GHz]{gigahertz}
\acrodef{Hz}[Hz]{hertz}
\acrodef{IF}[IF]{intermediate frequency}
\acrodef{IFFT}[IFFT]{inverse fast Fourier Transform}
\acrodef{KHz}[KHz]{kilohertz}
\acrodef{LO}[LO]{local oscillator}
\acrodef{LOS}[LOS]{line--of--sight}
\acrodef{MHz}[MHz]{megahertz}
\acrodef{MILTAL}[M\.{I}LTAL]{Millimeter Wave and Terahertz Technologies Research Laboratories}
\acrodef{MIMO}[MIMO]{multiple--input multiple--output}
\acrodef{mmWave}[mmWave]{millimeter wave}
\acrodef{NGWN}[NGWN]{next generation wireless network}
\acrodef{NLOS}[NLOS]{non line--of--sight}
\acrodef{OML}[OML]{Oleson Microwave Labs}
\acrodef{OLOS}[OLOS]{optical line--of--sight}
\acrodef{PNA}[PNA]{performance network analyzer}
\acrodef{QoS}[QoS]{quality of service}
\acrodef{RF}[RF]{radio frequency}
\acrodef{spar}[s--parameter]{scattering parameters}
\acrodef{subThz}[sub--\unit{}{THz}]{sub--terahertz}
\acrodef{TUBITAK}[T\"{U}B\.{I}TAK]{Scientific and Technological Research Council of Turkey}
\acrodef{USB}[USB]{universal serial bus}
\acrodef{VNA}[VNA]{vector network analyzer}
\acrodef{AoA}[AoA]{angle of arrival}
\acrodef{MLE}[MLE]{maximum likelihood estimation}

\ifCLASSINFOpdf
\else
\fi

\begin{document}
\title{Statistical Modeling of Propagation Channels for Terahertz Band\thanks{This paper is accepted for publication in 2017 IEEE Conference on Standards for Communications \& Networking (CSCN 2017).}} 
\author{\IEEEauthorblockN{Ali R{{\i}}za Ekti\IEEEauthorrefmark{1}\IEEEauthorrefmark{4}, Ali Boyac{{\i}}\IEEEauthorrefmark{3}, Altan Alparslan\IEEEauthorrefmark{1}\IEEEauthorrefmark{9}, {\.{I}}lhami \"{U}nal\IEEEauthorrefmark{2}, Serhan Yarkan\IEEEauthorrefmark{6},\\Ali G\"{o}r\c{c}in\IEEEauthorrefmark{1}\IEEEauthorrefmark{5}, H\"{u}seyin Arslan\IEEEauthorrefmark{8}, Murat Uysal\IEEEauthorrefmark{7}}

\IEEEauthorblockA{\IEEEauthorrefmark{1}Informatics and Information Security Research Center (B{\.{I}}LGEM), T{\"{U}}B{\.{I}}TAK, Kocaeli, Turkey}

\IEEEauthorblockA{\IEEEauthorrefmark{4}Department of Electrical--Electronics Engineering, Bal{{\i}}kesir University, Bal{{\i}}kesir, Turkey}

\IEEEauthorblockA{\IEEEauthorrefmark{3}Department of Electrical--Electronics Engineering, Istanbul Commerce University, {\.{I}}stanbul, Turkey}

\IEEEauthorblockA{\IEEEauthorrefmark{9}Department of Electrical and Electronics Engineering, Yeditepe University, {\.{I}}stanbul, Turkey}

\IEEEauthorblockA{\IEEEauthorrefmark{2}Millimeter Wave and Terahertz Technologies Research Laboratories (M{\.{I}}LTAL), T{\"{U}}B{\.{I}}TAK, Kocaeli, Turkey}

\IEEEauthorblockA{\IEEEauthorrefmark{6}Center for Applied Research on Informatics Technologies (CARIT), Istanbul Commerce University, {\.{I}}stanbul, Turkey}

\IEEEauthorblockA{\IEEEauthorrefmark{5}Faculty of Electronics and Communications Engineering, Y{{\i}}ld{{\i}}z Technical University, {\.{I}}stanbul, Turkey}

\IEEEauthorblockA{\IEEEauthorrefmark{8}Department of Electrical--Electronics Engineering, Medipol University, {\.{I}}stanbul, Turkey}

\IEEEauthorblockA{\IEEEauthorrefmark{7}Department of Electrical--Electronics Engineering, \"{O}zye\u{g}in University, {\.{I}}stanbul, Turkey\\ Emails: \texttt{arekti@balikesir.edu.tr,} \texttt{\{aboyaci, syarkan\}@ticaret.edu.tr,} \\ \texttt{altan.alparslan@std.yeditepe.edu.tr,} \texttt{ilhami.unal@tubitak.gov.tr,} \\\texttt{agorcin@yildiz.edu.tr,} \texttt{huseyinarslan@medipol.edu.tr,} \texttt{murat.uysal@ozyegin.edu.tr}}}

\maketitle

\begin{abstract}
Digital revolution and recent advances in telecommunications technology enable to design communication systems which operate within the regions close to the theoretical capacity limits. Ever--increasing demand for wireless communications and emerging numerous high--capacity services and applications mandate providers to employ more bandwidth--oriented solutions to meet the requirements. Trend and predictions point out that marketplace targets data rates around \unit{$10$}{Gbps} or even more within the upcoming decade. It is clear that such rates could only be achieved by employing more bandwidth with the state--of--the--art technology.  Considering the fact that bands in the range of \unit{$275$}{GHz}--\unit{$3000$}{GHz}, which are known as Terahertz (\unit{}{THz}) bands, are not allocated yet for specific active services around the globe, there is an enormous potential to achieve the desired data rates.
Although \unit{}{THz} bands look promising to achieve data rates on the order of several tens of \unit{}{Gbps}, realization of fully operational \unit{}{THz} communications systems obliges to carry out a multi--disciplinary effort including statistical propagation and channel characterizations, adaptive transceiver designs, reconfigurable platforms, advanced signal processing algorithms and techniques along with upper layer protocols equipped with various security and privacy levels. Therefore, in this study, several important statistical parameters for \ac{LOS} channels are measured. High resolution frequency domain measurements are carried out at single--sweep within a span of \unit{$60$}{GHz}. Impact of antenna misalignment under \ac{LOS} conditions is also investigated. By validating exponential decay of the received power in both time and frequency domain, path loss exponent is examined for different frequencies along with the frequency--dependent path loss phenomenon. Furthermore, impact of humidity is also tested under \ac{LOS} scenario. Measurement results are presented along with relevant discussions and future directions are provided as well.
\end{abstract}
%

\IEEEpeerreviewmaketitle
\acresetall

\section{Introduction}
Wireless data traffic has grown tremendously in the past years. According to the data analysis reports, total Internet traffic will reach \unit{$2.3$}{ZB} whose $66$\% will be generated by wireless devices \cite{cisco_2016}. It should be noted here that not only the volume of data increases exponentially, but also the number of users --and therefore-- devices escalate. Short--term predictions reveal that there will be seven billion people with seven trillion nodes/devices connected with each other via various forms of wireless communications technologies \cite{wwrf_2013_seven_billion_wireless}. Moreover, it is believed that marketplace targets data rates approaching multi--\unit{}{Gbps} even \unit{}{Tbps} within ten to fifteen years \cite{schneider_inf_milli_2015,akyildiz_phycom_2014}. Evidently, such rates could only be achieved by either extremely spectral efficient modulations or expanding the transmission bandwidth dramatically. Considering the fact that bands in the range of \unit{$275$}{GHz}--\unit{$3000$}{GHz}, which are known as Terahertz (\unit{}{THz}) bands, are not allocated yet for specific active services around the globe, there is an enormous potential to achieve the desired data rates.

Systems operating at \unit{}{THz} frequencies are attracting great interest and expected to meet the ever--increasing demand for high--capacity wireless communications as well as consumer expectations. Technological progress towards designing the electronic components operating at \unit{}{THz} frequencies will lead to a wide range of applications especially for short--range communications such as chip--to--chip communications, kiosk downloading, \ac{D2D} communications, and wireless backhauling \cite{kurner_inf_milli_2014,tajima_inf_milli_2016,akyildiz_phycom_2014}. Although \unit{}{THz} bands look promising to achieve data rates on the order of several tens of \unit{}{Gbps}, realization of fully operational \unit{}{THz} communications systems obliges to carry out a multi--disciplinary effort including statistical propagation and channel characterizations, adaptive transceiver designs (including both baseband and \ac{RF} front--end portions), reconfigurable platforms, advanced signal processing algorithms and techniques along with upper layer protocols equipped with various security and privacy levels. As in traditional wireless communications systems design process, realization of high--performance and reliable \unit{}{THz} communications systems should start with obtaining detailed knowledge about the statistical properties of the propagation channel. Next, these properties are incorporated into various channel characterizations and models. Upon verification and validation of the characterizations and models under different scenarios, system design stage is initiated at the end.

\subsection{Related Work}
Studies that focus on modeling the channel for \unit{}{THz} bands in the literature could be categorized in various ways. Measurement methodology; bandwidth; temporal, frequency, and spatial domain behaviors; and application--specific scenarios are some of them among others. Each and every measurement set concentrates on a specific scenario with some parameter changes. For instance, channel measurement results at \unit{$300$}{GHz} are described in \cite{priebe_tran_anten_prop_2011} for different scenarios focusing on path loss measurements under \ac{LOS}. 

Ray tracing is also employed in the literature to characterize the channel from the perspective of several parameters including regarding path loss, delay, angle of arrival, polarization, and angle of departure \cite{priebe_tran_wc_2013}. But ray tracing approach suffers from the following two problems: \begin{inparaenum}[(i)]\item{A minuscule change in the propagation environment mandates to redo the calculations. Each scenario should be treated separately. Therefore, no insight into fundamental statistics of the channel is provided.} \item{Calculations are computationally intensive and any increase in the area/volume causes the computation time to increase exponentially. }\end{inparaenum}

Bearing in mind that \unit{}{THz} bands allow vast amount of spectrum to be exploited, certain measurement and processing strategies should be devised due to the practical concerns. For instance, in \cite{khalid_icc_2016}, entire \unit{}{THz} region measurement data set is split into relatively smaller chunks and then post--processed. Note that such a ``divide--and--conquer'' approach is not safe, since there always exists a risk of creating artifacts during compilation. Spatial diversity through the use of \ac{MIMO} measurements are studied in literature as well. In \cite{khalid_wcl_2016}, a $2\times2$--\ac{MIMO} \ac{LOS} system operating at \unit{}{THz} region with \unit{$10$}{GHz} bandwidth is presented. However, there is a significant discrepancy between measurement data presented and the theoretical rates, which needs further research, verification, and validation. Apart from the traditional channel measurements, scenario--specific measurement results are presented in the literature as well. In \cite{kim_tran_wc_2016}, a $2$D geometrical propagation model for short--range \ac{D2D} desktop communication channels and multipath fading channels are considered. 

Having a brief overview of the measurement studies at hand,\footnote{Interested readers may refer to \cite[and references therein]{kurner_2012_towards} for further measurement results and campaigns present in the literature.} one could conclude that a systematic and comprehensive channel modeling strategy for \unit{}{THz} bands has not been established yet. Although \unit{}{THz} bands manifest many intrinsic propagation characteristics and mechanisms, \ac{LOS} state is preeminent among others because of the following reasons: First, \ac{LOS} is desired in \unit{}{THz} bands for high--performance operation. Second, \ac{LOS} presents the elementary propagation characteristics. In this regard, a detailed investigation of \ac{LOS} measurements should be the first step towards acquiring a systematic and comprehensive statistical channel model.

\subsection{Contributions}
Contribution of this study is three--fold considering the statistical channel characterization for \unit{}{THz} scenarios: \begin{inparaenum}[(i)]\item {To the best knowledge of authors', this work provides one of the first single--sweep THz measurement results within \unit{$240$}{GHz}--\unit{$300$}{GHz} band and relevant statistical analysis. } \item{Detailed statistical analyses of antenna--tilt measurement results under \ac{LOS} conditions within large--volume anechoic chamber are provided.} \item{In addition, impact of humidity is also considered under \ac{LOS} scenarios and relevant results are given.} \end{inparaenum}

\subsection{Paper Organization}

The remainder of this paper is structured as follows. System and signal model are depicted in \SEC{sec:statchar}. The measurement setup is presented in \SEC{sec:measurement}. Measurement  results and relevant discussions are given in \SEC{sec:meas_results}. Finally, conclusions are drawn in \SEC{sec:conclusion}.

\section{Statistical Characterization of Terahertz Channel}\label{sec:statchar}

\subsection{System and Signal Model}
\label{sec_system_and_signal_model}

\subsubsection{System Model}
\label{sec_system_model}

In this study, \ac{LOS} transmission scenario for \unit{}{THz} channels is investigated. As will be described in Section~\ref{sec:measurement}, a \ac{LOS} transmission scenario is established within an anechoic chamber along with very well--isolated setup through the use of absorbers.

In order to characterize the \unit{}{THz} channels, several factors and parameters such as channel geometry, relative motion, antenna directivity and radiation patterns along with environmental conditions (\textit{e. g.}, humidity, vapor, dust, etc.) and so on should all be taken into account accordingly. On the other hand, both analysis and transceiver design at further stages should be simplified as well. From this point of view, traditional linear, time--invariant channel model approach will be adopted.

\newcommand{\CB}[1]{\ensuremath{\left\{#1\right\}}}
\newcommand{\PP}[1]{\ensuremath{\left(#1\right)}}
\renewcommand\Re{\operatorname{Re}}

\subsubsection{Signal Model}
\label{sec_signal_model}

Any signal arriving at the receiver antenna at passband could be represented as:
\begin{equation}
r(t) = \Re{\CB{\PP{x_{I}(t)+jx_{Q}(t)}e^{j2\pi f_{c}t}}}
\end{equation}

\noindent where $j = \sqrt{-1}$; $\Re{\CB{\cdot}}$ is the real part of its complex input; $x_{I}(t)$ and $x_{Q}(t)$ represent the in--phase and quadrature components of the complex baseband equivalent of the transmitted signal, respectively; $f_{c}$ denotes the transmission frequency. Considering a static, general propagation environment, the received signal is modeled as a superposition of multiple copies of the transmitted signal with different delays and attenuation levels. Therefore, the channel at passband could be modeled with: $$h_{PB}(t) = \sum_{l = 0}^{L - 1} a_{l}\delta\PP{t - t_{l}}$$ where $h_{PB}(t)$ denotes the channel impulse response at baseband; $L$ is the number of resolvable multipath components; $a_{l}$ and $t_{l}$ represent attenuation factor and delay corresponding to $l$--th path, respectively; and $\delta(\cdot)$ represents the Dirac delta function. Complex baseband equivalent of the channel is given by $$h(t) = \sum_{l = 0}^{L - 1} a_{l}e^{-j2\pi f_{c}t_{l}}\delta\PP{t - t_{l}}$$

Static \ac{LOS} channels are a special sub--class of the multipath channels where $L = 1$. Therefore, the complex baseband equivalent channel can be described as:

\def\absorbtime{\ensuremath{a_{f}}}
\def\absorbfreq{\ensuremath{A(f)}}

\begin{equation}
\label{eq_channel_impulse_response_baseband}
h(t) = \absorbtime e^{j\theta}\delta(t - t_{0})
\end{equation}

\noindent where $\absorbtime$ represents the \ac{LOS} path amplitude with respect to the specular power affected by the environment geometry; $\theta$ is the phase; $t_{0}$ denotes the propagation delay as a deterministic value of $t_{0} = d/c$ with $d$ denoting the transmitter--receiver separation and $c$ is the speed of light (\unit{$3\times 10^{8}$}{m/s}). It is important to note here that in case directional antennas used, which is the common practice in \unit{}{THz} communications, the impact of misalignment of the antennae; frequency--dependent loss; and frequency dispersion index could all be absorbed by the term, $\absorbtime$.

In the literature, stochastic description of multipath components in a static environment is generally considered to be superposition of both specular and diffused components such that \cite{yarkan_2008_LOS}:
\begin{equation}
\begin{split}
\label{eq_LOS_analysis}
m_{l} 	&= a_{l}e^{-j2\pi f_{c}t_{l}}\\ 
		&= s_{l} + d_{l}
\end{split}
\end{equation} 

\noindent In \eqref{eq_LOS_analysis},

\begin{equation}
s_{l} = \sigma_{s_{l}}e^{\PP{j2\pi f_{c}\cos{\PP{\theta_{l}}} + \phi_{l}}}
\end{equation}

\noindent and
\begin{equation}
d_{l} = \sigma_{d_{l}}\frac{1}{\sqrt{M_{l}}}\sum_{m=1}^{M_{l}}b_{m}e^{\PP{j2\pi f_{c}\cos{\PP{\theta_{m}}} + \phi_{m}}}
\end{equation}

\noindent where $\sigma_{s_{l}}$ denotes the magnitude; $\theta_{l}$ is the \ac{AoA}; and $\phi_{l}$ is the phase of the specular component, respectively. Similarly, $\sigma_{d_{l}}$ is the magnitude of the diffused component; $M_{l}$ represents the number; $b_{m}$ is the amplitude; $\theta_{m}$ is the \ac{AoA}; and $\phi_{m}$ is the phase of the incoming waves forming the diffused component, respectively. Without loss of generality, both $\sigma_{s_{l}}$ and $\sigma_{d_{l}}$ could be assumed to be unity under idealized conditions. Therefore, $a_{l}$ could theoretically be represented with Rayleigh distributed in the absence of \ac{LOS} component under uniform \ac{AoA} assumption. In case \ac{LOS} is present, Rice and Nakagami distributions are employed to describe the first--order statistics of $a_{l}$. 

Because \ac{LOS} is a special case among all other general wireless propagation scenarios, both large-- and small--scale fading mechanisms exhibit intrinsic characteristics. As will be discussed in Section~\ref{sec:measurement}, fully isolated measurement setup with absorbers in an anechoic chamber guarantees the \ac{LOS} transmission. This implies that the losses introduced by propagation channel are limited to distance--dependent path loss, possible antenna misalignment(s), and equipment imperfections along with non--ideal behaviors due to operating around/above ideal regions. In this regard, signal model could be deduced down to a direct path experiencing a distant--dependent path loss without shadowing with a possible misalignment. This implies that the contribution of path loss to the term $\absorbtime$ in \eqref{eq_channel_impulse_response_baseband} could be given by:
\begin{equation}
PL = PL_{0} + 10\pathloss\log_{10}{(d)} + M
\end{equation}

\noindent where $PL_{0}$ denotes the path loss at a reference distance in the antenna far--field; $n$ is the path loss exponent whose value depends on the propagation environment and conditions within; $d$ is the transmitter--receiver separation; and $M$ is the random variable representing the impact of misalignment as a random antenna gain. In the literature, due to its tractability and plausibility, misalignment is considered only from the transmitter or receiver antenna perspective as a zero--mean Gaussian random variable with certain standard deviation, $\sigma_{M}$, representing the misalignment.

\section{Terahertz Measurement Experiment Setup}\label{sec:measurement}

\subsection{Description of Measurement Setup}

\begin{figure}[!t]
	\centering
	\includegraphics[width=3.35in,height=2in]{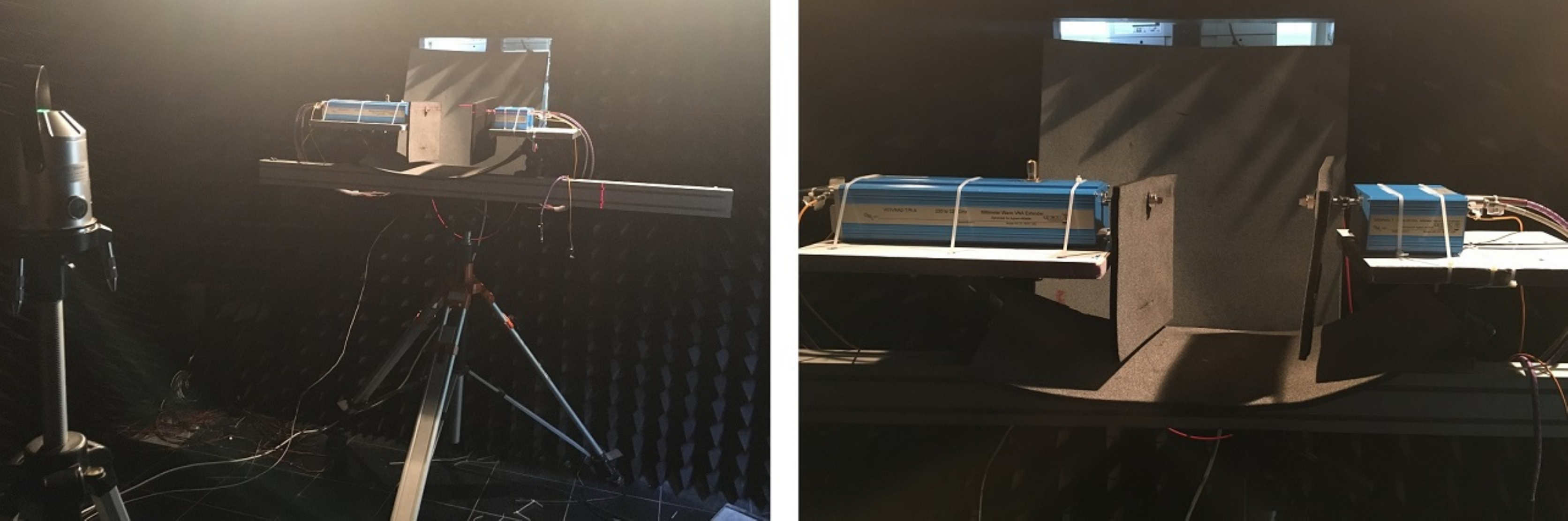}
	\caption{Measurement setup in anechoic chamber from two different angles. Note that \ac{LOS} conditions are emulated by well--isolated measurement equipment in the anechoic chamber.}
	\label{fig_meas_env}
\end{figure}

We constructed an experimental measurement setup in the \ac{MILTAL} at the \ac{TUBITAK} in Gebze, Turkey and it can be seen in Figure~\ref{fig_meas_env}. The dimensions of the anechoic chamber used in the measurements are \unit{$7$}{m}$\times$\unit{$3$}{m}$\times$\unit{$4$}{m} (length$\times$width$\times$height).

\begin{figure}[!t]
\centering
\includegraphics[width=3.35in,height=1.75in]{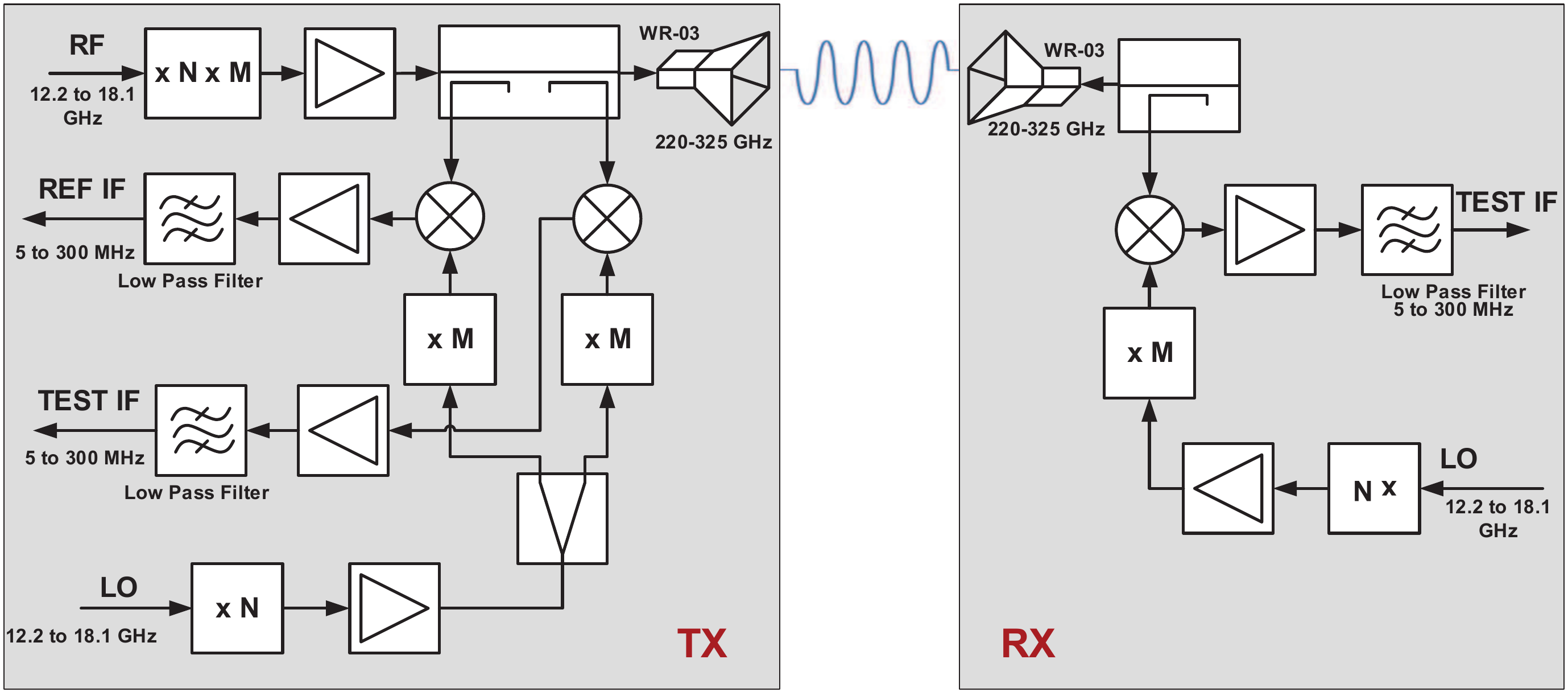}
\caption{Block diagram of the measurement setup.}
\label{fig_sim}
\end{figure}

The measurement setup consists of four major parts: Agilent \ac{PNA} \ac{VNA} E$8361$A, \ac{OML} V$03$VNA$2$--T and V$03$VNA$2$--T/R--A millimeter wave extender modules and N$5260$A extender controller. Since the upper limit of the \ac{VNA} is \unit{$67$}{GHz}, the \unit{$220$}{GHz} to \unit{$325$}{GHz} extender modules are attached to the \ac{VNA} using the controller to measure channel characteristics at \unit{}{THz} frequencies. V$03$VNA$2$--T/R--A extender module contains multipliers ($\times18$) that extends \unit{$12.2$}{GHz} to \unit{$18.1$}{GHz} \ac{RF} input signal to \unit{$220$}{GHz} to \unit{$325$}{GHz} frequency range. Test \ac{IF} and reference \ac{IF} signals for \ac{VNA} are obtained by using downconversion mixers before transmiting. After passing through the channel, the received signal is downconverted at the receiver module, V$03$VNA$2$--T, by using downconversion mixers and the resulting test \ac{IF} (\unit{$5$}{MHz} to \unit{$300$}{MHz}) is fed back to the \ac{VNA}. Difference in the transmitted and received signal is analyzed to find channel characteristics. The corresponding block diagram of our setup is shown in Figure~\ref{fig_sim}.

The \unit{$220$}{GHz} to \unit{$325$}{GHz} source modules include balanced multipliers, which are driven by an extended band WR--$10$ multiplier chain with a WR--$03$ waveguide output interface. The output source match of the modules is typically \unit{$9$}{dB}. The \ac{RF} drive signal may be either \ac{CW} or swept frequency. The required \ac{RF} or \ac{LO} signal power to operate \ac{OML} modules is \unit{$+10$}{dBm}. The dynamic range is typically \unit{$75$}{dB} (min. \unit{$60$}{dB}) and the WR--$03$ waveguide output power of the V$03$VNA$2$--T/R--A is around \unit{$-23$}{dBm}. The magnitude and phase stability of the extender modules are \unit{$\pm 0.4$}{dB} and \unit{$\pm 8$}{\si{\degree}}, respectively.

In this study, \unit{$240$}{GHz} to \unit{$300$}{GHz} is specially used to get better performance from the extender modules, in terms of magnitude and phase stability. Channel transfer function is acquired by recording \ac{spar} using the measurement setup shown in Figure~\ref{fig_meas_env} and Figure~\ref{fig_sim}. For channel modeling, calibration is performed with direct interconnection of the transmitting and receiving extender module's waveguide ports. All the later measurements are taken with standard gain horn antenna, with the gain of \unit{$24.8$}{dBi} at the center frequency, attached at both the transmitter and the receiver. Full \unit{$60$}{GHz} band measurements are recorded with $4096$ points averaging and \unit{$100$}{Hz} \ac{IF} \ac{BW}. These parameters significantly reduce noise floor and improves dynamic range. 

\subsection{Measurement Methodology}\label{subsec:measmethod}

\ac{VNA} measurements are performed for the frequency interval of \unit{$240$}{GHz} to \unit{$300$}{GHz}. This interval is measured using standard gain horn antennas which are attached to \ac{OML} extenders. Each spectral measurement is represented with $4096$ equally spaced frequency points (data points) within the interval specified by the \ac{VNA}. Therefore, a spectral resolution of \unit{$14.648$}{MHz} is obtained. The measurement system including connectors and cables is calibrated to remove the impairment caused by the components. The calibration data is saved both into the internal memory of the Agilent \ac{PNA} \ac{VNA} E$8361$A and an external \ac{USB} storage device. Then, the calibration data is removed from the measurement by using its internal memory and provided $S_{21}$ parameter in complex number format with real and imaginary parts for each data point.

\section{Measurement Results}\label{sec:meas_results}

\begin{figure}[!t]
\centering
\includegraphics[width=3.45in,height=2.35in]{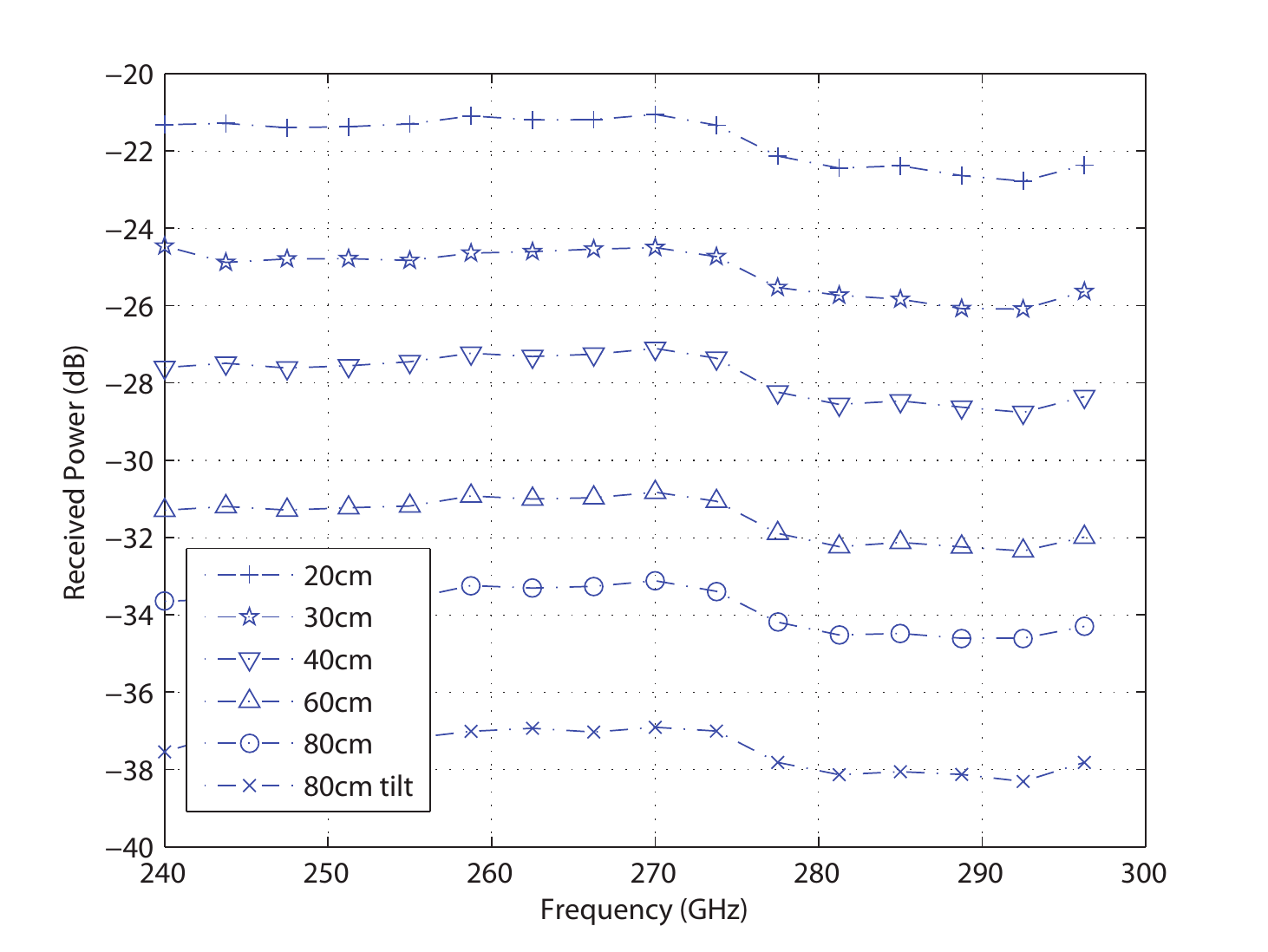}
\caption{Averaged channel frequency responses in logarithmic scale for various \ac{LOS} scenarios including the impact of antenna misalignment via antenna tilt.}
\label{fig_cfr_v1_log}
\end{figure}

As stated before, channel magnitude response for \unit{}{THz} channels is an important qualitative tool. Especially wideband measurements reveal different mechanisms whose effects are obscured in traditional channel characterizations due to relatively narrower band measurements. Frequency--dependent loss is one of them. In this regard, channel magnitude response could be used to analyze the frequency dependency of the loss in frequency domain very easily. In Figure~\ref{fig_cfr_v1_log}, averaged magnitude responses are given. However, first it is appropriate to evaluate the distance--dependent path loss. Based on the measurement data, the path loss coefficients for specific frequencies are given in Table~\ref{tbl_path_loss_exponents_for_specific}. Overall mean path loss exponent is found to be $n = 1.9704$ based on $4096$--point resolution with a variance of $\approx 0.003$ by taking into account entire \unit{$60$}{GHz} span. This result is in conformity with the  \ac{LOS} argument and with the measurement results reported in the literature \cite{kim_2015_300GHz_propagation_desktop,sawada_2016_300GHz_path_loss}. Assuming that the path loss exponent distributed normally, the maximum likelihood estimate of the parameters of the distribution match with the mean perfectly, whereas $\sigma_{\text{MLE}_{n}}^{2} = 0.0591$.

Another important aspect of wideband measurements is to be able to observe the impact of frequency--dependent path loss. Note that frequency dependency stems from the antenna  not from the transmission path itself in . This dependency is clearly seen in Figure~\ref{fig_cfr_v1_log} between \unit{$0.27$--$0.29$}{THz} as a slight collapse.

\begin{table}[]
\centering
\caption{Path Loss Exponents for Specific Frequencies Along with Overall Statistics}
\label{tbl_path_loss_exponents_for_specific}
\begin{tabular}{@{}lccccccc@{}}
\toprule
Freq. (\unit{}{GHz}) 	& $240$   & $250$   & $260$   & $270$   & $280$   & $290$   & $300$   \\
\midrule
PL. Exp. ($n$)      		& $2.02$ & $2.04$ & $1.96$ & $1.90$ & $1.96$ & $1.94$ & $2.01$\\
\midrule
\multicolumn{8}{c}{Overall Statistics}\\
\midrule
Mean ($n$)      		& \multicolumn{7}{c}{$1.9704$, based on $4096$ points}\\
\midrule
Variance ($n$)		& \multicolumn{7}{c}{$0.0035$, based on $4096$ points}\\
\bottomrule
\end{tabular}
\end{table}

\begin{figure}[!t]
\centering
\includegraphics[width=3.45in,height=2.35in]{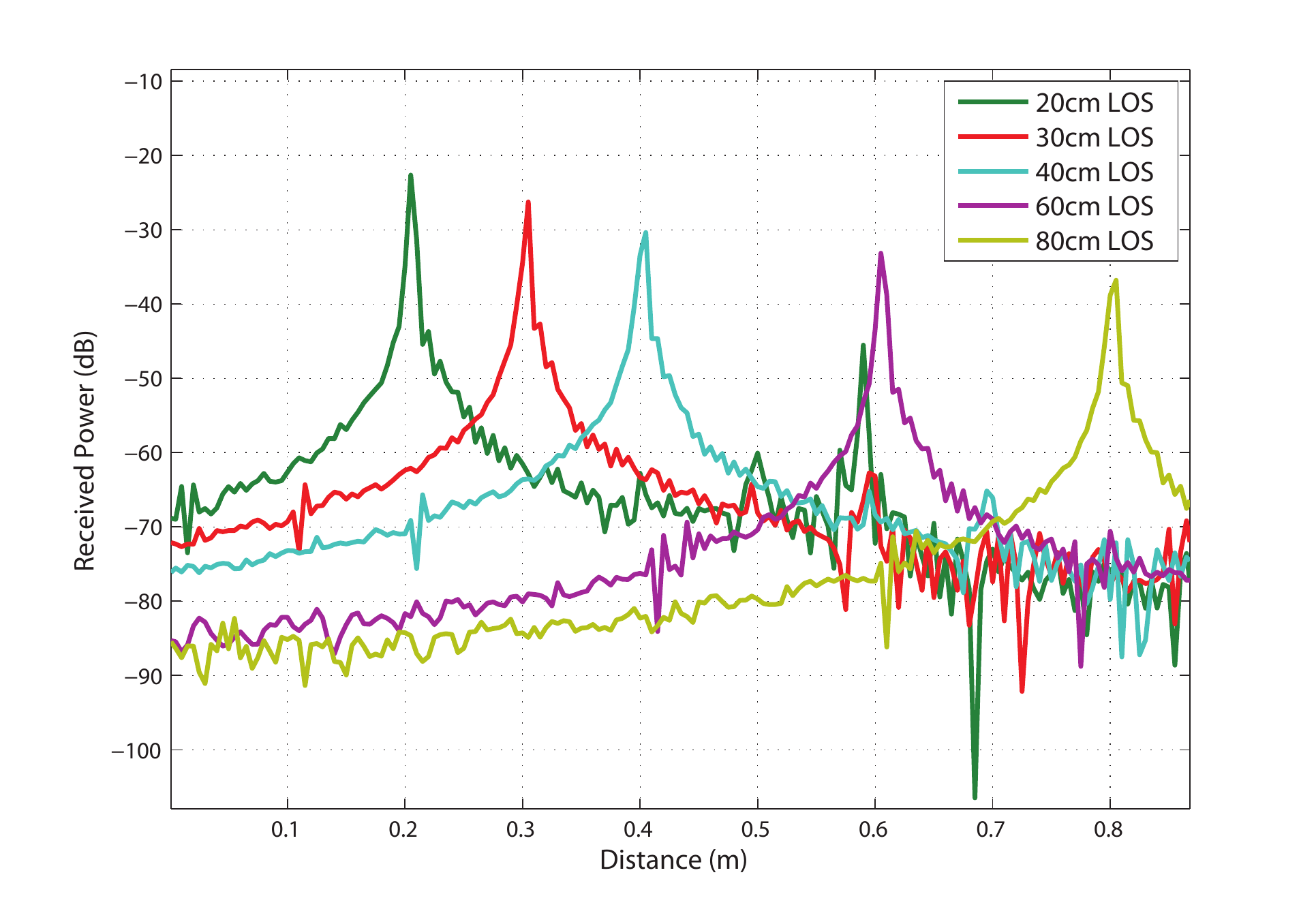}
\caption{First arriving paths in temporal domain for different transmitter--receiver separations under \ac{LOS}. The horizontal axis is given in terms of transmitter--separation to validate the experimental setup values.}
\label{fig_cir_v1_log}
\end{figure}

\begin{figure}[!t]
	\centering
	\includegraphics[width=3.45in,height=2.35in]{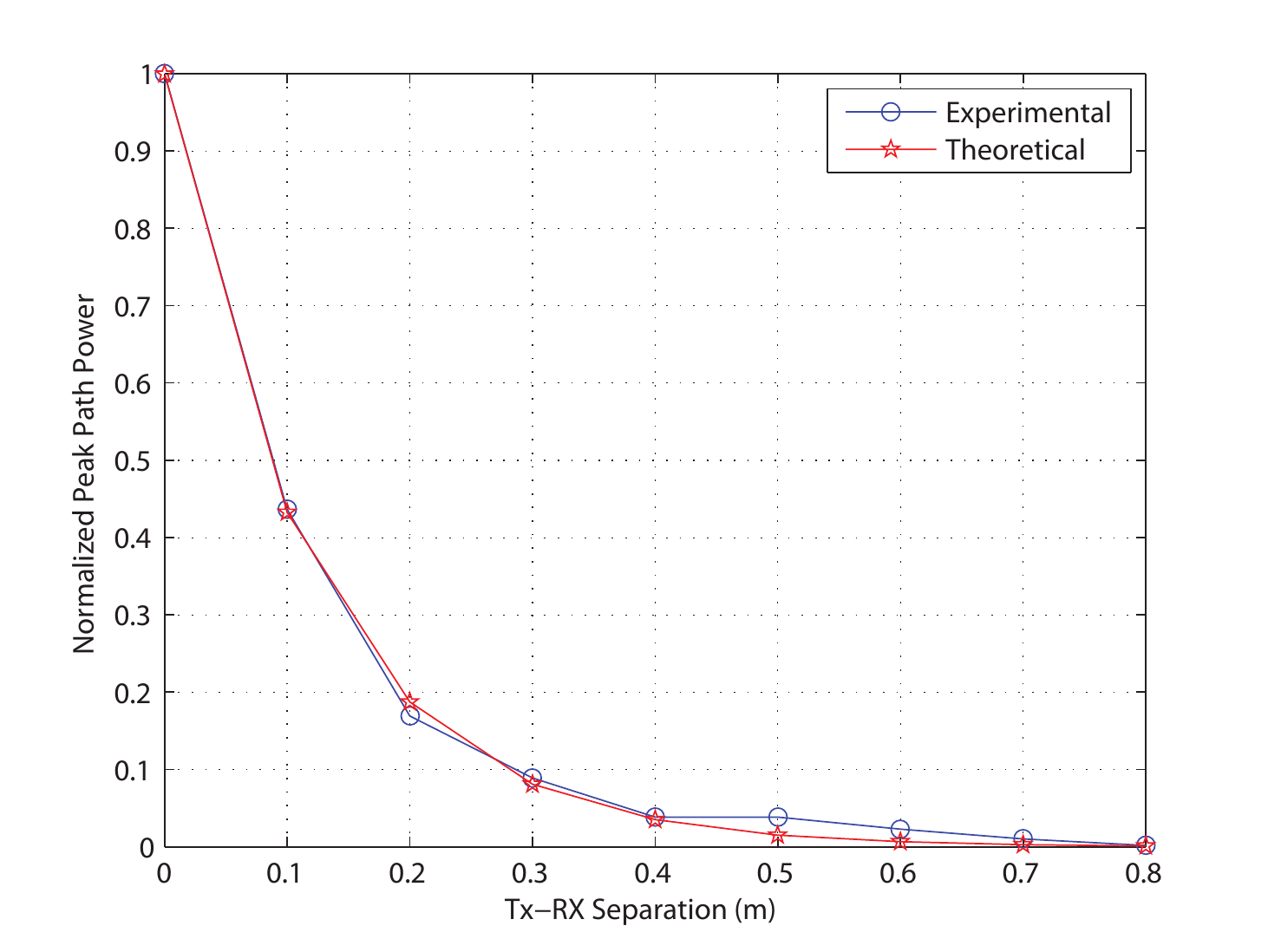}
	\caption{Normalized measured peak \ac{LOS} path power levels along with corresponding exponential decay function whose parameter is obtained by \ac{MLE}.}
	\label{fig_cir_v1_linear}
\end{figure}

In order to validate the frequency domain results, time domain analysis is carried out as well. By applying \ac{IFFT} operation, time domain data are obtained. Raw time data are plotted in Figure~\ref{fig_cir_v1_log}. In Figure~\ref{fig_cir_v1_log}, it is seen that the delay $t_{0}$ is given in terms of corresponding distance matching with the experimental setup values. After validating the post--processing stage by delay--distance match, raw time data are post--processed further by removing the propagation delay and normalizing the power with respect to the reference power. Results of this procedure are given in Figure~\ref{fig_cir_v1_linear}. As a cross--check for the tabulated results given in Table~\ref{tbl_path_loss_exponents_for_specific}, exponential decay in the received power is validated through the use of the impulse responses obtained. Here, it is assumed that the underlying distribution is of exponential form with $f(x;\lambda) = \lambda e^{-\lambda x}$ where $\lambda$ is the parameter to be estimated for candidate exponentially distributed random variable $X$ where $x\in\mathbb{R}^{+}$. In addition, assuming that the underlying distribution of the power levels of paths observed/measured is the same exponential distribution with the parameter $\lambda$ and that all $N$ measurements are independent of each other, the likelihood function, $f(x_{1}, x_{2}, \cdots, x_{N};\lambda)$, for the observed/measured data becomes $$\mathcal{L}(\lambda) = \prod_{k = 1}^{N}f(x_{k};\lambda) = \lambda^{N} e^{-\lambda\PP{\sum_{k = 1}^{N}x_{k}}}$$

\noindent Applying logarithm on each side reads $\ln{\PP{\mathcal{L}(\lambda)}} = N\ln{\PP{\lambda}} - \lambda\sum_{k = 1}^{N}x_{k}$. Finally, differentiating the $\log$--likelihood function $\ln{\PP{\mathcal{L}(\lambda)}}$ with respect to $\lambda$ and equating it to zero yields $N/\lambda - \sum_{k = 1}^{N}x_{k} = 0$. Therefore, the maximum likelihood estimate of the parameter $\lambda$ is obtained by $$\hat{\lambda} = \frac{\sum_{k = 1}^{N}x_{k}}{N}$$ Both the measured data and corresponding maximum likelihood estimated theoretical curve are plotted in Figure~\ref{fig_cir_v1_linear}. Almost perfect match between the measured and \ac{MLE}--driven values could be recognized easily in the figure.

\begin{figure}[!t]
\centering
\includegraphics[width=3.45in,height=2.35in]{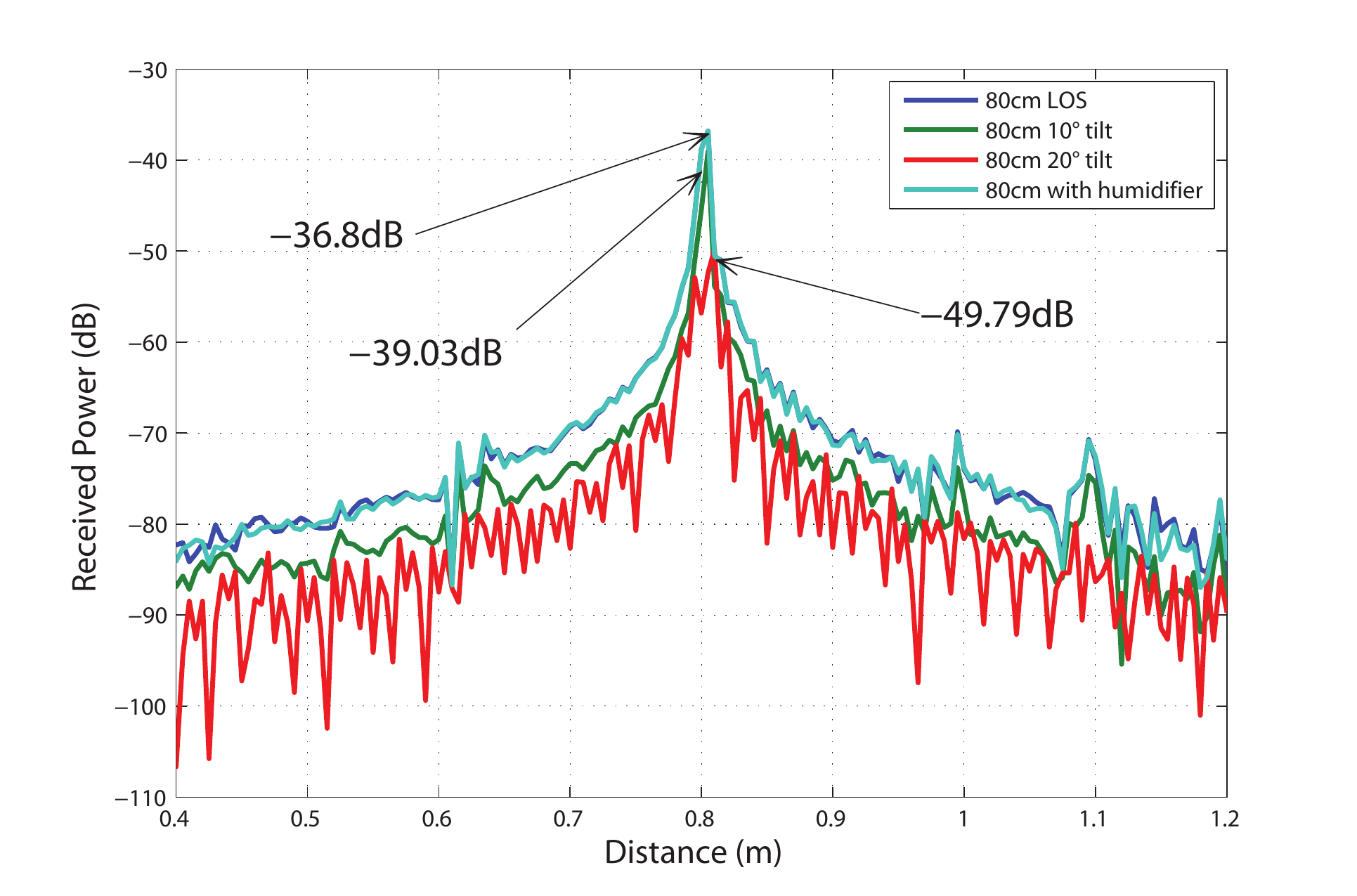}
\caption{Measured channel impulse responses at \unit{$80$}{cm} in logarithmic scale with several antenna tilts along with a separate measurement in the presence of dense humidity.}
\label{fig_cir_80cm_degree_humidifer}
\end{figure}

As discussed in Section~\ref{sec_system_and_signal_model}, antenna misalignment is an important problem for \unit{}{THz} communication channels. In this regard, how antenna misalignment affects the received power is evaluated in time domain. The results are given in Figure~\ref{fig_cir_80cm_degree_humidifer}. As can be seen from the figure, antenna misalignment causes a significant drop in the peak power corresponding to \ac{LOS} path. Measurement results demonstrate that \unit{$10$}{$\si{\degree}$} tilt leads to \unit{$\approx 2.3$}{dB}, whereas \unit{$20$}{$\si{\degree}$} tilt causes \unit{$\approx 13$}{dB} degradation for the peak power. Note that degradation in power level due to antenna misalignment could also be verified in frequency domain as plotted in Figure~\ref{fig_cfr_v1_log}. 

One of the main factors that affects the propagation behavior of \unit{}{THz} band signals is the presence of water vapor within the environment. It is known that water vapor molecules introduce not only attenuation but also colored noise \cite{jornet_2010_THz_channel_capacity,jornet_2011_THz_channel_modeling}. In this regard, the impact of the humidity is also investigated for \unit{}{THz} communication channels. Channel impulse response obtained under a humid environment is appended into Figure~\ref{fig_cir_80cm_degree_humidifer} as well. It is seen that peak power level of the \ac{LOS} path does not exhibit a significant drop in parallel with the measurement results reported in the literature \cite{siles_anten_prop_mag_2015}.

\section{Concluding Remarks and Future Directions}\label{sec:conclusion}
Deploying communication systems operating within \unit{}{THz} bands is considered to be an alternative strategy to meet the ever--increasing data rate demands along with escalating number of devices subscribing wireless networks. Due to the technical limitations and propagation loss concerns, \unit{}{THz} bands have not attracted a significant attention up until the last couple of years. However, with the technological advances, it is possible to migrate up to \unit{}{THz} region. Nevertheless, a successful and reliable communication system relies heavily on well--established propagation channel models and appropriate transceiver designs. 

In this study, single--sweep band measurement data for \unit{$240$}{GHz}--\unit{$3000$}{GHz} band are collected in frequency domain with a very high resolution within an anechoic chamber along with a very well isolated setup to emulate the  propagation. Behavior of the channel within a \unit{$60$}{GHz} span (\textit{i.e.}, \unit{$240$}{GHz}--\unit{$300$}{GHz} interval) is captured at once. In addition, high--resolution measurement data are collected so that finer temporal details are obtained to help design reliable transceiver systems including antenna misalignment problem. Since  scenario provides a theoretical borderline, collected data could be used to validate other results obtained in different measurement campaigns.

Practical applications for \unit{}{THz} communications are generally envisioned to operate in short--range such as infotainment systems. This implies that locations and spatial orientations of \unit{}{THz} devices could be random especially in residential scenarios.In this regard, \ac{NLOS} behavior of the channels especially for indoor applications should be examined in detail. Even though there are several studies focusing on \ac{LOS} and \ac{NLOS} in the literature, more results are required in order to have a more comprehensive understanding of the channels. Finally, how mobility affects the channel behavior should be studied under different conditions with various measurement campaigns.
\section*{Acknowledgment}
The authors would like to thank Mr. Mustafa K\i{}l\i{}\c{c} from M\.{I}LTAL, T{\"{U}}B{\.{I}}TAK for his help during the measurements.

\bibliographystyle{IEEEtran}

\end{document}